\documentclass[twocolumn,amsmath,amssymb,nofootinbib,a4paper]{revtex4} 
\usepackage[utf8]{inputenc}
\usepackage{graphicx} 
\usepackage{amsmath}
\usepackage{amsfonts,amsbsy}
\usepackage{amssymb}
\usepackage{subdepth}
\usepackage{bbold}
\usepackage{nicefrac}
\usepackage{xcolor}
\definecolor{lcolor}{rgb}{0.5,0,0}
\definecolor{citcolor}{rgb}{0,0.3,0.0}
\usepackage[breaklinks,colorlinks,urlcolor=blue,citecolor=citcolor,linkcolor=lcolor]{hyperref}

\newcommand{\nc}{{N_\mathrm{c}}}

\newcommand{\qs}{Q_\mathrm{s}}

\newcommand{\nr}[1]{(\ref{#1})} 

\newcommand{\fig}{Fig.~}
\newcommand{\figs}{Figs.~}
\newcommand{\eq}{Eq.~}

\begin{document}

\title{Plasmon mass scale in classical nonequilibrium gauge theory}

\author{T. Lappi}
\email{tuomas.v.v.lappi@jyu.fi}
\affiliation{
Department of Physics, %
 P.O. Box 35, 40014 University of Jyv\"askyl\"a, Finland
}
\affiliation{
Helsinki Institute of Physics, P.O. Box 64, 00014 University of Helsinki,
Finland
}

\author{J. Peuron}
\email{jarkko.t.peuron@student.jyu.fi}
\affiliation{
Department of Physics, %
 P.O. Box 35, 40014 University of Jyv\"askyl\"a, Finland
}

\begin{abstract}
Classical lattice Yang-Mills calculations provide a good way to understand different nonequilibrium phenomena in nonperturbatively overoccupied systems. Above the Debye scale the classical theory can be matched smoothly to kinetic theory. The aim of this work is to study the limits of this quasiparticle picture by determining the plasmon mass in classical real-time Yang-Mills theory on a lattice in three spatial dimensions. We compare three methods to determine the plasmon mass: a hard thermal loop expression in terms of the particle distribution, an effective dispersion relation constructed from fields and their time derivatives, and the measurement of oscillations between electric and magnetic field modes after artificially introducing a homogeneous  color electric field. We find that a version of the dispersion relation that uses electric fields and their time derivatives agrees with the other methods within  50\%.
\end{abstract}

\maketitle

\section{Introduction}
The classical field approximation is commonly used to study time-dependent phenomena in gauge-field theory. In the weak coupling limit $g\ll1$, it is justified for modes that have a nonperturbatively  high occupation number $f\sim 1/g^2$ for gluonic states. This happens in thermal systems for relatively infrared modes for which the Bose-Einstein occupation number increases as $f\sim T/k$. Here  the classical approximation can be used for the ``electric''  modes with $p\sim gT$, while the dominant modes $p\sim T$ must be described as fully quantum fields, or as classical particles.

In high-energy collisions of hadrons, on the other hand, the physics of gluon saturation leads to the emergence of a semihard  dominant transverse momentum scale $\qs$ at which the occupation numbers are large. In this case, plasma instabilities~\cite{Mrowczynski:1994xv,Mrowczynski:1996vh,Mrowczynski:2004kv} have been argued to dominate the early stage of isotropization towards a thermal plasma~\cite{Arnold:2003rq,Romatschke:2003ms,Romatschke:2004jh,Arnold:2004ti,Arnold:2004ih}. This picture has been confirmed both in Boltzmann-Vlasov~\cite{Nara:2005fr,Dumitru:2005gp,Bodeker:2007fw,
Rebhan:2008uj,Attems:2012js} and also in purely classical Yang-Mills (CYM) simulations~\cite{Romatschke:2005pm,Romatschke:2006nk,Berges:2008zt}. More recently classical field simulations have also been used to understand the creation of CP-violating fluctuations in the early stages of a heavy-ion collision~\cite{Mace:2016svc,Mueller:2016ven}. The growth rate of the plasma instabilities is parametrically given by the Debye or plasmon mass scale.

The Debye or plasmon mass in this context is a well-defined quantitative concept in hard-thermal-loop (HTL) perturbation theory. In the HTL case there is a clear separation of scales at weak coupling, with most of the energy of the system residing in modes with $p\sim T$, where the occupation numbers are of order 1. The power counting is very different in the overoccupied case considered in heavy-ion collisions, where the energy resides in modes $p\sim \qs$ with occupation number $f\sim 1/g^2$. In the CYM calculation the coupling constant scales out completely, and the small value of $g$ does not introduce a scale separation between the dominant modes $p\sim \qs$ and the Debye or plasmon scale. It is clear from the previous works (e.g.~\cite{Krasnitz:2000gz,Romatschke:2005pm,Romatschke:2006nk,Mace:2016svc,Mueller:2016ven}), however, that the Debye or plasmon scale nevertheless also exists in the classical theory. Indeed the behavior of the classical fields seems to be remarkably well described by a kinetic theory description~\cite{Kurkela:2012hp} in terms of quasiparticle degrees of freedom. The purpose of this paper is to study this picture in more detail to understand to what extent an overoccupied classical gauge theory system can be understood in a quasiparticle picture.

In particular, the aim of this paper is to develop  and compare numerical methods to determine the plasmon mass in a strongly occupied, nonequilibrium, dynamical  system of gauge fields. In HTL theory it is  of the same order as the Debye mass (these two quantities differ  by a constant factor). However, what we are studying here are time-dependent oscillations of gauge fields (plasmons), not the Debye screening of static color charges. Therefore we use henceforth the term ``plasmon mass'' which more accurately describes the aim here.
 We will compare systematically three methods of extracting  a plasmon mass: extracting from a HTL-approximation formula in terms of an integral over the quasiparticle number distribution, extracting from the oscillation frequency of a homogenous chromoelectric field, and extracting from comparing correlators of Coulomb-gauge  fields and their time derivatives, which we refer to as the ``dispersion relation'' (DR) method.  In this paper we will focus on a three-dimensional isotropic system, for which the comparison to a thermal one is most straightforward.   We plan to return to strongly anisotropic systems exhibiting plasma instabilities in future work, taking advantage of the methods developed here.

We will first discuss the initial setup of our real-time lattice calculation in Sec.~\ref{sec:method} and the three methods for determining the plasmon mass in Sec.~\ref{sec:omegapl}. We then test the dependence of the results on the infrared and ultraviolet cutoffs present in the lattice calculation in Sec.~\ref{sec:cutoff} and then on the physical parameters of our system, the time and occupation number, in Sec.~\ref{sec:results}, before concluding in Sec.~\ref{sec:conc}.

\begin{figure}[t!]
\centerline{\includegraphics[width=0.48\textwidth]{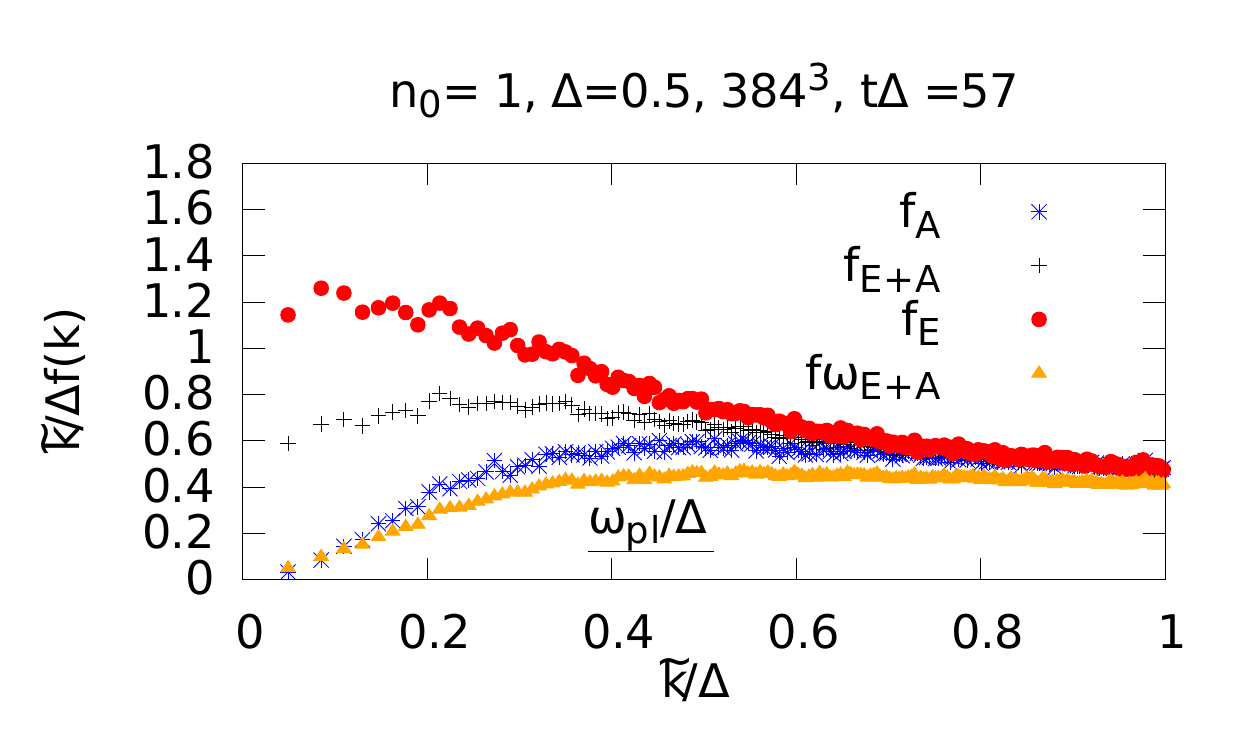}}
 \caption{Different methods of defining particle number distribution. Here $f\omega_{E+A}$ stands for a particle distribution extracted using Eq. (\ref{eq:discF}) with a massive dispersion relation, while the others are extracted assuming a massless dispersion relation. The two other distributions ($f_A$ and $f_E$) are obtained by assuming equal distribution of electric and magnetic energy in each mode, which allows us to assume that two terms contribute equally and use only the other multiplied by 2. The rough location of the plasmon mass scale has also been indicated. }
\label{fig:fmethods}
\end{figure}

\section{Numerical method and initial conditions}\label{sec:method}

\subsection{Equations of motion in the temporal gauge}
All numerical simulations in this paper are done by using the SU(2) gauge group for numerical convenience. We do not expect a qualitative difference in the dynamics between SU(2) and SU(3)~\cite{Berges:2008zt,Ipp:2010uy}.
The equations of motion used are given by the standard Wilson action on a three-dimensional lattice 
\begin{align}
\mathcal{S} & =  -\beta_0 \sum_{x,i} \left( \frac{1}{N} \mathfrak{Re}\mathrm{Tr}\left(\underline{\Box}_x^{0,i}\right) - 	1 \right) \\ & +  \beta_s \sum_{x,i<j} \left( \frac{1}{N} \mathfrak{Re}\mathrm{Tr}\left(\underline{\Box}_x^{i,j}\right) -1 \right), 
\label{eq:wilsonaction}
\end{align}
where $\beta_0 = \frac{2 N \gamma}{g^2},$ $\beta_s = \frac{2 N}{g^2 \gamma},$  $\gamma = \frac{a_s}{a_t}$ and $U_{x,i}$ are the link matrices defined as 
\begin{equation}
U_{x,i} = \exp{\left(i a_s g A_i\left(x\right) \right)}.
\label{eq:linkdef}
\end{equation}
 The plaquette is defined as $\underline{\Box}_x^{i,j} \equiv U_{x,i} U_{x+i,j} U^\dagger_{x+j,i} U^\dagger_{x,j }$ and can be related to the exponential of the field strength tensor.  The spatial lattice spacing is $a_s$ and the temporal one  $a_t.$ We use the standard normalization for the generators of SU(2), i.e. $\mathrm{Tr}\left(t^a t^b\right) = \dfrac{1}{2}\delta^{ab}.$ 

\begin{figure}[t!]
\centerline{\includegraphics[width=0.48\textwidth]{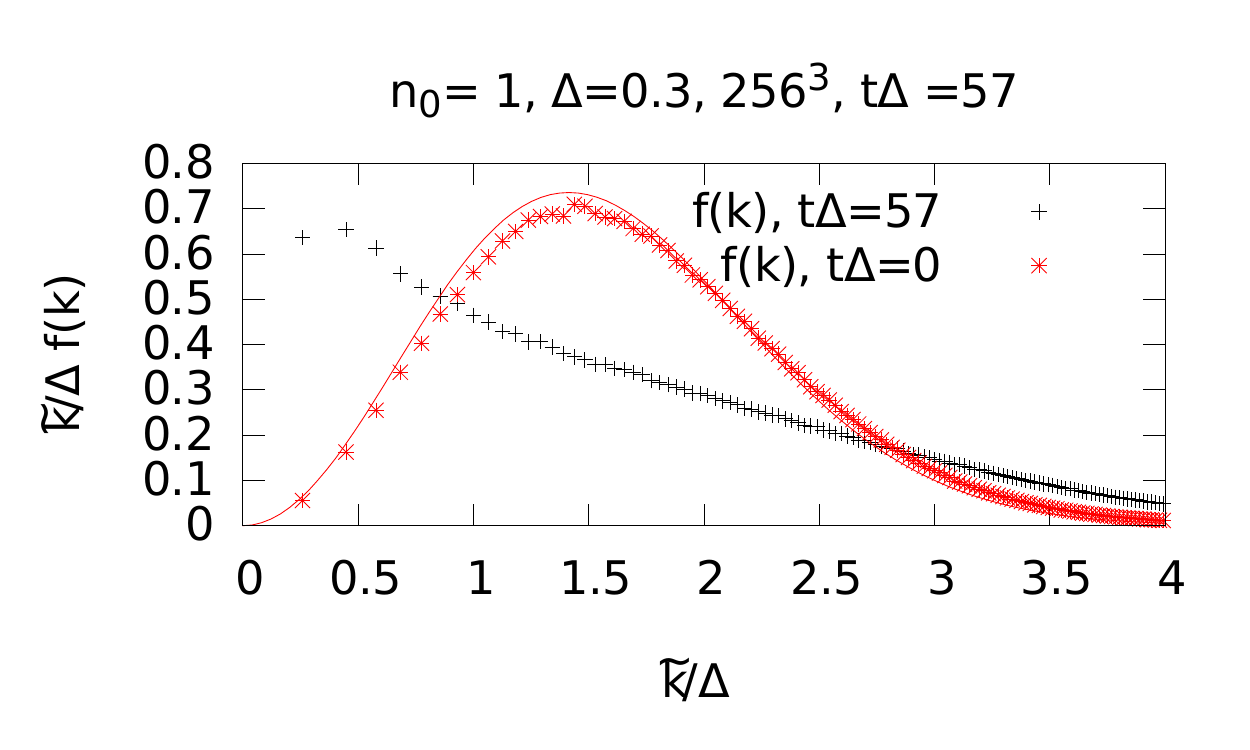}}
 \caption{Particle number distribution at the initial time and at $t \Delta=57$, showing also the analytical form of the initial distribution.}
\label{fig:f}
\end{figure}

When we extract the physical fields from our simulations we use the following definitions:
\begin{equation}
E_i^a(x) = \dfrac{2}{a_s a_t g} \mathfrak{Im}\mathrm{Tr}(t^a\underline{\Box}_x^{i,0})
\end{equation}
\begin{equation}
B_i^a(x) = -\dfrac{ \varepsilon_{ijk}}{a_s^2 g} \mathfrak{Im}\mathrm{Tr} \left( t^a \underline{\Box}_x^{j,k} \right)
\end{equation}
\begin{equation}
 F_{\mu \nu}^a(x) = \dfrac{2}{ a_\mu a_\nu g} \mathfrak{Im}\mathrm{Tr} \left( t^a \underline{\Box}_x^{\mu,\nu} \right)
 \end{equation}
\begin{equation}
 A_\mu^a(x) = \dfrac{2}{a_\mu g } \mathfrak{Im} \mathrm{Tr} \left( t^a U_{x,\mu} \right),
\label{eq:Aextraction} 
 \end{equation}
 where $a_\mu$ refers to the lattice spacing in the $\mu$ direction.
One can easily verify that the rhs approaches the continuum counterpart of the lhs when we take lattice spacings to zero. 
Varying the action (\ref{eq:wilsonaction}) with respect to the spatial links gives the equations of motion for the electric field \begin{align}
\label{eq:Eevolution}
E_j (t , x)  & =  E_j(t-a_t , x) + \dfrac{a_t}{2 i a_s^3 g} \sum_k \Bigg(  \underline{\Box}_x^{j,k}   - \underline{\Box}_x^{k,j} \nonumber \\ &- \dfrac {\mathbb{1}}{N} \mathrm{Tr} \left(  \underline{\Box}_x^{j,k}   - \underline{\Box}_x^{k,j} \right) 
+ \overline{\Box}_x^{j,k} - \left(\overline{\Box}_x^{j,k}\right)^\dagger  \nonumber \\ & - \dfrac{\mathbb{1}}{N} \mathrm{Tr} \left( \overline{\Box}_x^{j,k} - \left(\overline{\Box}_x^{j,k}\right)^\dagger \right) \Bigg),  
\end{align}
where $\overline{\Box}_x^{j,k} = U_{x,j} U_{x+j-k,k}^\dagger U_{x-k,j}^\dagger U_{x-k,k}.$ 
The links can be updated on the next time step by using the definition of the electric field on the lattice and the following decomposition (which holds for a SU(2) matrix)
\begin{equation}
\underline{\Box}_x^{i,0} = \sqrt{1 - \left(\dfrac{a_s a_t g}{2} E_a \right)^2}\mathbb{1} + ia_s a_t g E^a t^a.
\end{equation}
The temporal plaquette in the temporal gauge is just a product of link matrices at two different time steps, so we can easily solve for the link at the next time step.

Varying the action (\ref{eq:wilsonaction}) with respect to temporal links gives a nondynamical constraint, which is the non-Abelian analogue of the Gauss's law in classical electrodynamics
\begin{equation}
\sum_j \left( E_j(x) - U_{x-j, j }^\dagger E_j(x-j) U_{x-j, j } \right) = 0.
\label{eq:gauss}
\end{equation}
This constraint is also conserved by the discretized equations of motion.

\subsection{Quasiparticle distribution}\label{sec:partdist}

There is no unique way to determine a quasiparticle distribution from a given classical field configuration (see also the discussion in Ref.~\cite{Kurkela:2012hp}). Here we start by gauge transforming the fields to Coulomb gauge in order to eliminate the gauge degrees of freedom. We utilize a Fourier-accelerated algorithm for the gauge fixing \cite{Davies:1987vs}, and we have also checked that increasing the gauge fixing precision does not change the observed quasiparticle spectrum.
If it is valid to describe the system as a collection of weakly interacting quasiparticles, the energy density of the system is given in terms  of this quasiparticle spectrum  by 
\begin{equation}
\epsilon = 2 \left(\nc^2-1\right)\int \dfrac{\mathrm{d}^3 k}{\left(2 \pi \right)^3} \omega\left(k\right) f\left(k\right).
\end{equation}
On the other hand, the total energy of the system is given by the Yang-Mills Hamiltonian 
\begin{equation}
\mathcal{H} = \int \mathrm{d}^3x \mathrm{Tr}\left( E_i E^i + B_i B^i\right).
\end{equation}
If we now keep only the quadratic terms in the fields and equate these two we find an expression for the quasiparticle spectrum 
\begin{equation} \label{eq:discF}
f\left(k\right) = \dfrac{1}{2} \dfrac{1}{2 \left(N_c^2-1 \right)} \dfrac{1}{V} \left(\dfrac{\left| E_C\left(k\right)\right|^2}{\omega\left(k\right)} + \dfrac{k^2}{\omega\left(k\right)} \left|A_C\left(k\right)\right|^2\right).
\end{equation}
Here the dispersion relation is given by $\omega\left(k\right)$. Unless otherwise stated, we  assume a massless linear dispersion relation while extracting the quasiparticle spectrum and refer to this (massless) quasiparticle spectrum as $f$. We can also use a massive dispersion relation (with a plasmon mass extracted as discussed later in Sec.~\ref{subsec:DR}), for which we will use the notation $f_\omega.$ 
Because the mass increases $\omega(k)$ in the denominator, this reduces the estimate for the infrared occupancies. 
The data obtained is then averaged, if large statistical fluctuations are present, to smoothen the fluctuations and then interpolated using cubic splines.

If we assume that electric and magnetic modes carry an equal amount of energy (as is the case in a free theory in a time-averaged sense), we can replace the sum of the electric or magnetic field energies by only one of them multiplied by 2. A comparison of these different methods is shown in Fig.~\ref{fig:fmethods}. We find that the different expressions become inequivalent below the Debye scale. 

Unless otherwise stated,  we use \eq\nr{eq:discF} with a massless dispersion relation when referring to the quasiparticle spectrum, as we do not need to assume an equal distribution of energy between electric and magnetic modes. One must, however, keep in mind the significant ambiguity from the precise definition of the number distribution in what follows.

\subsection{Initial conditions}
We sample our initial condition from the following distribution
\begin{equation}
\left< A_i^a\left(k\right) A_j^b\left(p\right)\right> = \dfrac{V n_0}{g^2 \Delta } \exp{\left(\dfrac{-k^2}{2 \Delta^2}\right)} \delta_{ij} \delta^{ab} \dfrac{\delta^{\left(3\right)}\left(k+p\right) \left(2 \pi \right)^3}{V}.
\label{eq:IC}
\end{equation}
Here $V$ is the lattice volume and $\Delta$ is the dominant momentum scale. Although the momentum distribution is not exactly the same as in the early stages of a heavy-ion collision (in particular we only consider isotropic systems here),  $\Delta$ should be thought of as analogous to the saturation scale $\qs$ \cite{Gyulassy:2004zy}. Our initial condition contains purely magnetic energy, which is the most straighforward way to satisfy Gauss's law. Otherwise the main reason for choosing the Gaussian form \nr{eq:IC} is that it has a very clear dominant momentum scale $\Delta$, and behaves well both in the ultraviolet and infrared. We will measure e.g. momenta and times relative to this scale. The quasiparticle spectrum corresponding to the initial condition is
\begin{equation}
f\left(k, t=0 \right) = \frac{n_0}{g^2} \dfrac{k}{\Delta} \exp{\left( \dfrac{-k^2}{2 \Delta^2}\right)}.
\label{eq:initdist}
\end{equation}
Thus the normalization parameter $n_0$ controls the typical occupation number at the momentum scale $\Delta$, and should be $\gtrsim 1$ for the classical approximation to be valid for describing these degrees of freedom. 
This initial momentum distribution is the same as that used in Refs.~\cite{Berges:2007re,Berges:2012ev}. It is also close to the theta function used in Refs~\cite{Berges:2013eia,Berges:2014bba} in the sense of being very strongly cut off in the UV. For a more realistic initial condition for heavy-ion collisions see e.g.~\cite{Lappi:2011ju,Berges:2012cj}.
The particle number distribution at the initial condition and later at a typical time scale used in our simulation ($t\Delta=57$) are shown in Fig.~\ref{fig:f}. The initial deviation from the analytical curve is most likely caused by the fact, that the initial links are obtained by using equation (\ref{eq:linkdef}), and the gauge fields are extracted from the links using (\ref{eq:Aextraction}) which are  inverse operations only in the limit $a_s\to 0$. 

We do not expect the late-time behavior of the system to be strongly influenced by our choice of initial condition, as can be seen from Fig.~\ref{fig:f}, since the time evolution will rapidly alter the initial occupation number distribution.  All results in this work have been obtained from one simulation unless otherwise stated. Within one simulation we average over many momentum modes, which generates plenty of statistics especially on larger lattice sizes.

\section{Methods for extracting the plasmon mass}\label{sec:omegapl}

\subsection{Uniform electric field}

The first method we use is to introduce a spatially homogenous chromoelectric field on top of the original field and then measure the oscillation frequency between the electric and magnetic field energy, as described in \cite{Kurkela:2012hp}. The spatially homogenous field corresponds to introducing a plasma oscillation in the zero mode. The drawback of this method is that it is destructive; i.e., adding the homogenous field will perturb the values given by the other methods later, rendering them useless in the remaining simulation. It is also computationally expensive compared to the other methods: one has to run the simulation for possibly thousands of time steps to obtain a single estimate for the dispersion relation at a specific momentum. Introducing the spatially uniform electric field also explicitly breaks Gauss's law (\ref{eq:gauss}), and one has to restore it by hand, which we do here  using the algorithm described in Ref.~\cite{Moore:1996qs}.

\begin{figure}
\centerline{\includegraphics[width=0.48\textwidth]{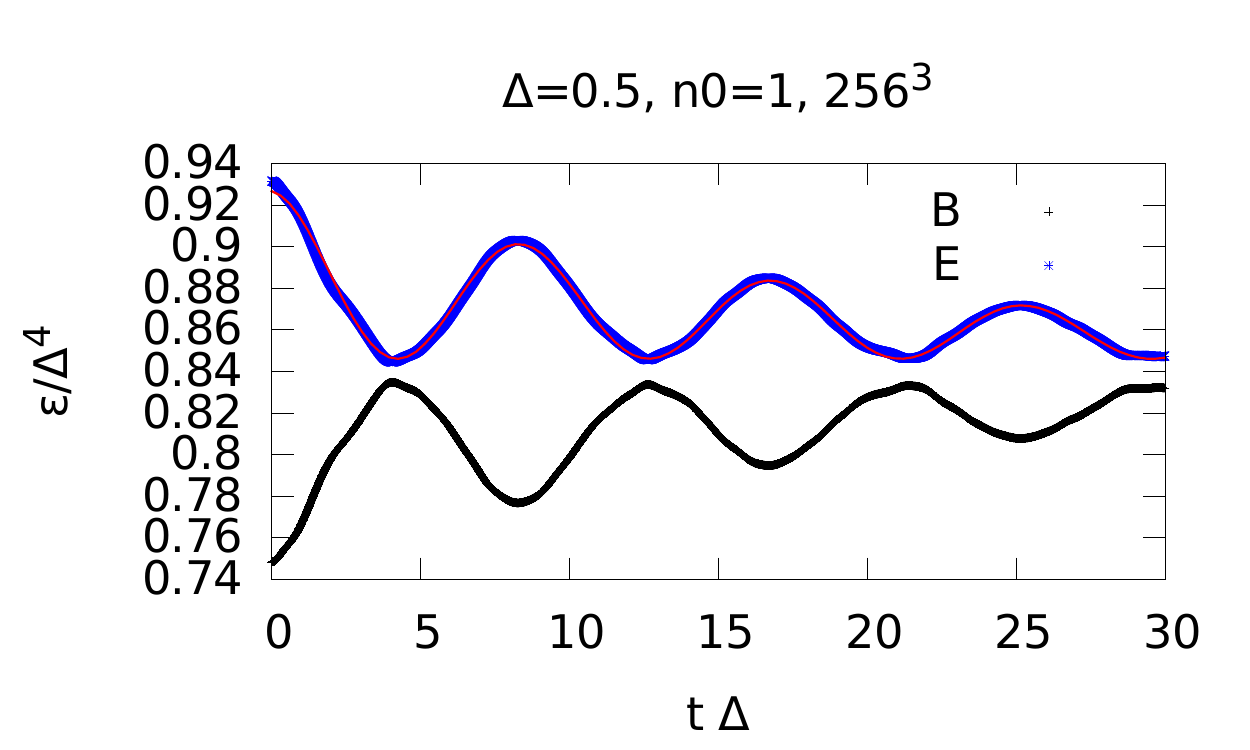}}
 \caption{Oscillating dimensionless energy density in the chromoelectric (E) and - magnetic fields (B) after the addition of the homogenous chromoelectric field. The plasmon mass given by the fit is $\nicefrac{\omega^2}{\Delta^2} \approx 0.14$ and the corresponding damping rate is $\nicefrac{\gamma^2}{\Delta^2} \approx 0.002,$ which is consistent with the damping rate extracted using the dispersion relation method. In order to simplify the fitting procedure, we have removed all energy data prior to adding the spatially uniform electric field. Thus we start counting the time from this point on. The physical time ($N_{\text{steps}}a_t\Delta$) elapsed before introducing the uniform electric field is 60 here. }
 \label{fig:ueosc}
\end{figure}

\begin{figure}
\centerline{\includegraphics[width=0.48\textwidth]{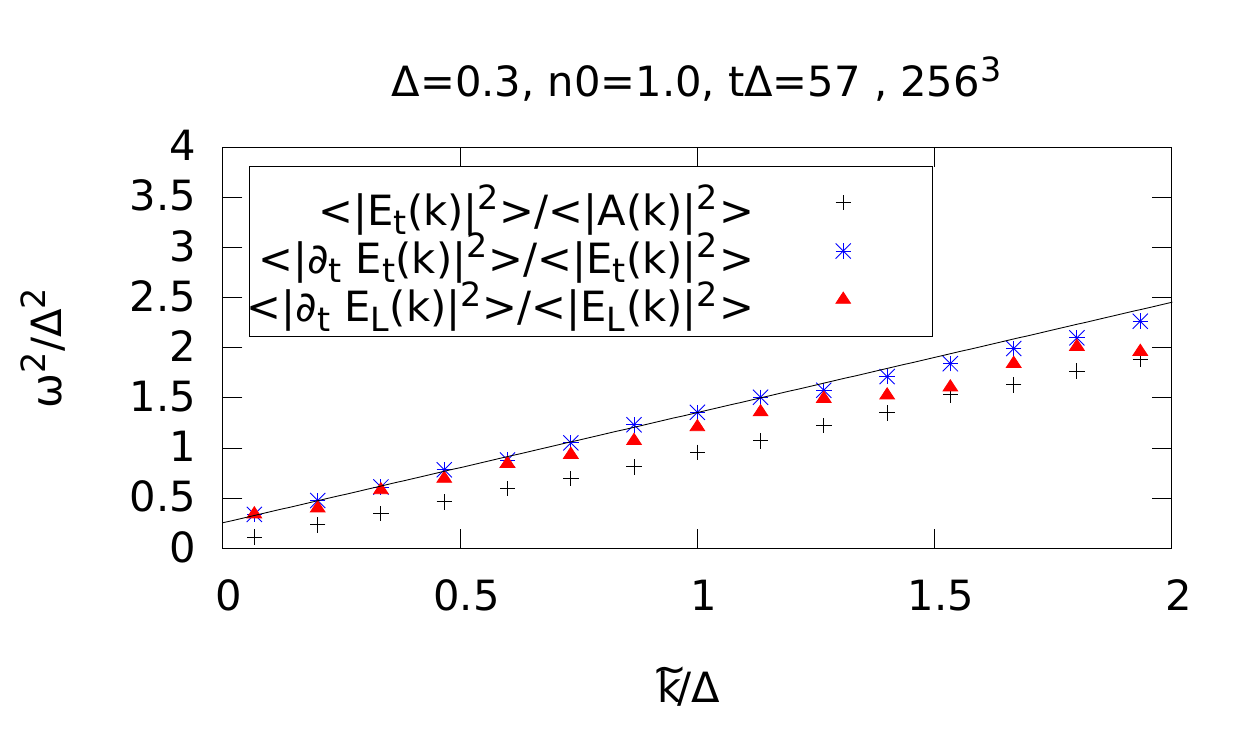}}
 \caption{Different estimates for the dispersion relation. The black dots show the dispersion relation as given by eq. (\ref{eq:karvalakkidispersiorelaatio}), but the longitudinal modes have been eliminated from the electric field.  The blue and red data points show the transverse and longitudinal dispersion relations respectively. The result we get when we extrapolate the dispersion relation of the transverse plasmons to zero momentum is $\nicefrac{\omega^2}{\Delta^2} = 0.256.$}
 \label{fig:DR}
\end{figure}

Figure \ref{fig:ueosc} shows a typical oscillation that takes place after we have introduced the uniform electric field. The magnitude of the added uniform electric field has been chosen in such a way that it increases the total energy of the system by approximately 10~\%. The reason for this is that we do not want to perturb the system too much, and on the other hand one needs to introduce a sufficiently strong electric field in order to get a signal clean enough for the extraction of the plasmon mass. As we can see the oscillation is well parametrized by a fit of the form $a+ b\cos^2\left(\omega t\right)e^{-\gamma t}.$ The ability to extract a damping rate simultaneously with the mass from this fit is an additional advantage of this method.

We have also studied the sensitivity of our results on the amount of energy added into the system when introducing the uniform electric field. When the introduced energy lies between 3 \% and 30 \% of the total energy of the system the change in the observed $\nicefrac{\omega_{pl}^2}{\Delta^2}$ is approximately within 5 \%. For the damping rate the change is approximately 25 \%. As we will find, the damping rate is roughly 2 orders of magnitude smaller than the plasmon mass, so this larger uncertainty will not affect our determination of the plasmon mass.

\subsection{Dispersion relation} \label{subsec:DR}
In order to extract the dispersion relation we gauge transform the fields into the Coulomb gauge. The standard approach has been to use 
\begin{equation}
\omega^2\left(k\right) = \dfrac{\left<\left|E_i^a\left(k\right)\right|^2 \right>}{\left< \left|A_i^a\left(k\right)\right|^2 \right>},
\label{eq:karvalakkidispersiorelaatio}
\end{equation}
as in \cite{Krasnitz:2000gz} in (2+1)-dimensional gauge theory. This has also been used in three-dimensional gauge theory, for example \cite{Berges:2012ev}.
As we can see from \fig \ref{fig:DR} it turns out this gives a very small value for the plasmon mass, and thus it does not agree with other methods at least in the three-dimensional case. 

Because we are imposing the Coulomb gauge, we are  eliminating the longitudinal component of the gauge potential $A_i$. However, there are still longitudinally polarized oscillations present in the system. Their magnetic part is hidden in the nonlinear terms in the gauge potential, but they are present in the electric field correlator. Thus if one wants to have same number of degrees of freedom in the numerator and denominator, one should project out the longitudinal components of the electric field to study the purely transverse dispersion relation. 

\begin{figure}[t!]
\centerline{\includegraphics[width=0.48\textwidth]{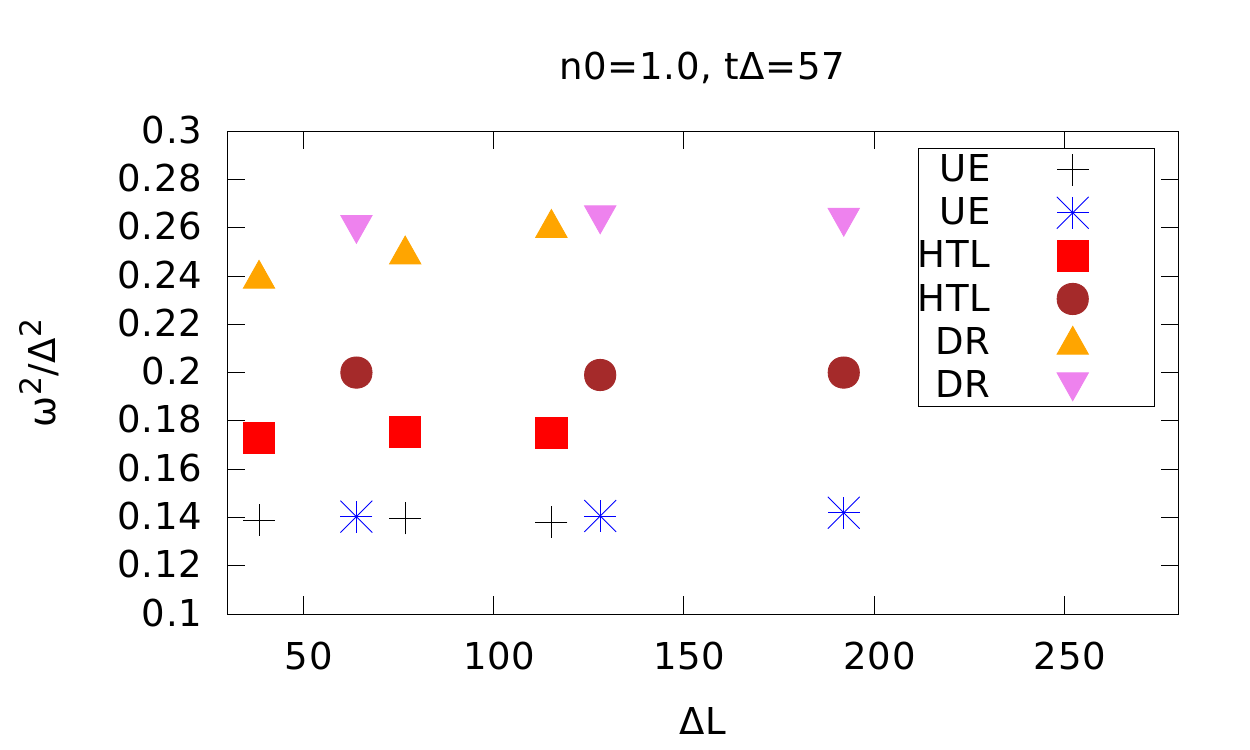}}
 \caption{Dependence of the plasmon mass on the system size (i.e. the infrared cutoff of the calculation). DR stands for dispersion relation, HTL for hard-thermal-loop resummed approximation and UE for the uniform electric field. The different values for $a_s \Delta$ correspond to different UV cutoffs. When we keep $a_s \Delta$ fixed, we find that our results do not depend on the infrared cutoff. In the key the data points are arranged pairwise in such a way, that the upper one corresponds to $a_s \Delta=0.3$ and the lower one to $a_s \Delta =0.5.$ }
 \label{fig:ircutoff}
\end{figure}

To do this, we can  separate the longitudinal and transverse modes by using the standard projection operators
\begin{align}
P_T^{ij} & = \delta^{ij} - \dfrac{\tilde{p}^i \tilde{p}^j}{\tilde{p}^2} \\
P_L^{ij} & = \dfrac{\tilde{p}^i \tilde{p}^j}{\tilde{p}^2},
\end{align}
where $\tilde{p}$ is given by
\begin{equation}
\tilde{p}_i = \dfrac{2}{a_s}\sin\left(\dfrac{p_i a_s}{2}\right).
\end{equation}
Here $p_i$ is the lattice momentum (here) defined as $p_i = \dfrac{\pi n_i}{L_i},$ with $n_i$ integers from 0 to $L_i-1,$ and $L_i$ number of points on the lattice in the $i$ direction.

On the lattice one must be careful with the projection mentioned above. 
The electric field resides on the link and is naturally centered at $x+\nicefrac{\hat{\imath}}{2}.$ Its ``physical'' Fourier transform should thus be defined as 
\begin{align}
E_i\left(k\right) &= \int \mathrm{d}^nx E\left(\vec{x}\right) e^{-i \vec{k}\cdot \left(\vec{x}+\nicefrac{\hat{\imath}}{2}\right)} \\ &= e^{i k \cdot \nicefrac{\hat{\imath}}{2}}\int \mathrm{d}^nx E\left(\vec{x}\right) e^{-i \vec{k}\cdot \vec{x}} ,
\end{align}
This additional phase will not contribute when we take the absolute value of the field squared as in \eq\nr{eq:karvalakkidispersiorelaatio}, but when decomposing the field into  transverse and longitudinal projections these phase factors need to be taken into account.

Equation \nr{eq:karvalakkidispersiorelaatio} yields an estimate for the plasmon mass that is far from what we would expect based on the other methods.  Armed with the transverse and longitudinal projectors, we can use another  estimate for the dispersion relation using the time derivative of the electric field,
\begin{equation}
\omega^2_{T,L}\left(k\right) = \dfrac{\left<\left|\dot{E}_{T,L,i}^a\left(k\right)\right|^2 \right>}{\left< \left|E_{T,L,i}^a\left(k\right)\right|^2 \right>}, 
\label{eq:omega2}
\end{equation}
where the dot stands for the time derivative.  Assuming a wave with time dependence $e^{i\omega t - \gamma t},$ this expression actually gives us $\omega^2 + \gamma^2.$ 
We can also extract the damping rate and dispersion relation separately from the data using the following expressions
\begin{equation}
\omega^2\left(p\right) = \dfrac{\left<\left(\mathfrak{Re}\dot{E}\right)^2\right>}{\left<\left(\mathfrak{Re}E\right)^2\right>} - \left|\dfrac{\left<\mathfrak{Re}E \mathfrak{Re}\dot{E}\right>}{\left<\left(\mathfrak{Re}E\right)^2\right>}\right|^2
\label{eq:omega2b}
\end{equation}
\begin{equation}
\gamma^2\left(p\right) =  \left|\dfrac{\left<\mathfrak{Re}E \mathfrak{Re}\dot{E}\right>}{\left<\left(\mathfrak{Re}E\right)^2\right>}\right|^2.
\label{eq:gamma2b}
\end{equation}
It turns out (given sufficient statistics) that the expression \nr{eq:omega2b}  converges to approximately the same value as  \eq(\ref{eq:omega2}). We observe numerically that the damping rate given by \nr{eq:gamma2b} is negligible compared to the plasmon mass, and thus we can safely use \eq(\ref{eq:omega2}) to estimate the frequency $\omega^2_{T,L}\left(k\right)$. In the following we denote as the ``dispersion relation'' plasmon mass the result of a fit of the form $\omega^2 = \omega_{pl}^2 + a k^2$ (with two free parameters $\omega_{pl}^2$ and $a$) to the numerical dispersion relation. When performing this fit, the values which we get for the slope are close to unity, which is the value we would physically expect.

\begin{figure}[t!]
\centerline{\includegraphics[width=0.48\textwidth]{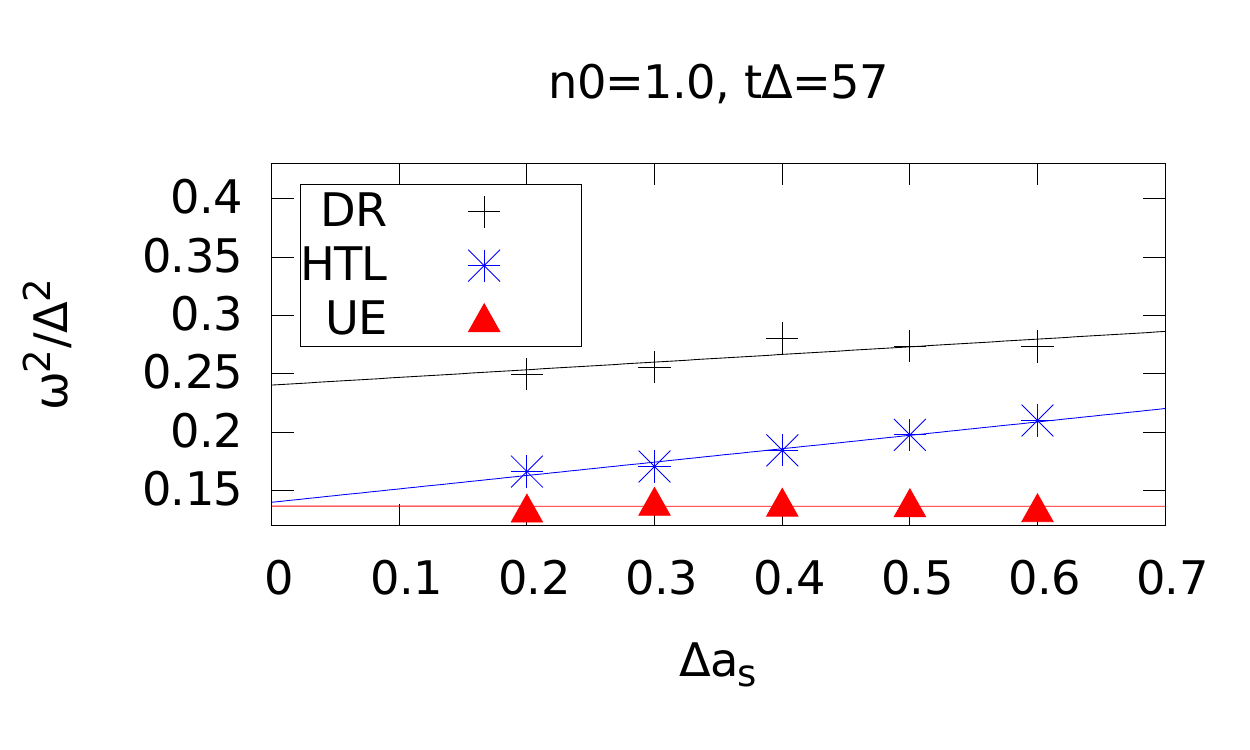}}
 \caption{Dependence of the plasmon mass on the ultraviolet cutoff. Straight lines show a linear extrapolation to $a_s=0$.  }
 \label{fig:uvcutoff}
\end{figure}

\begin{figure}[t!]
\centerline{\includegraphics[width=0.48\textwidth]{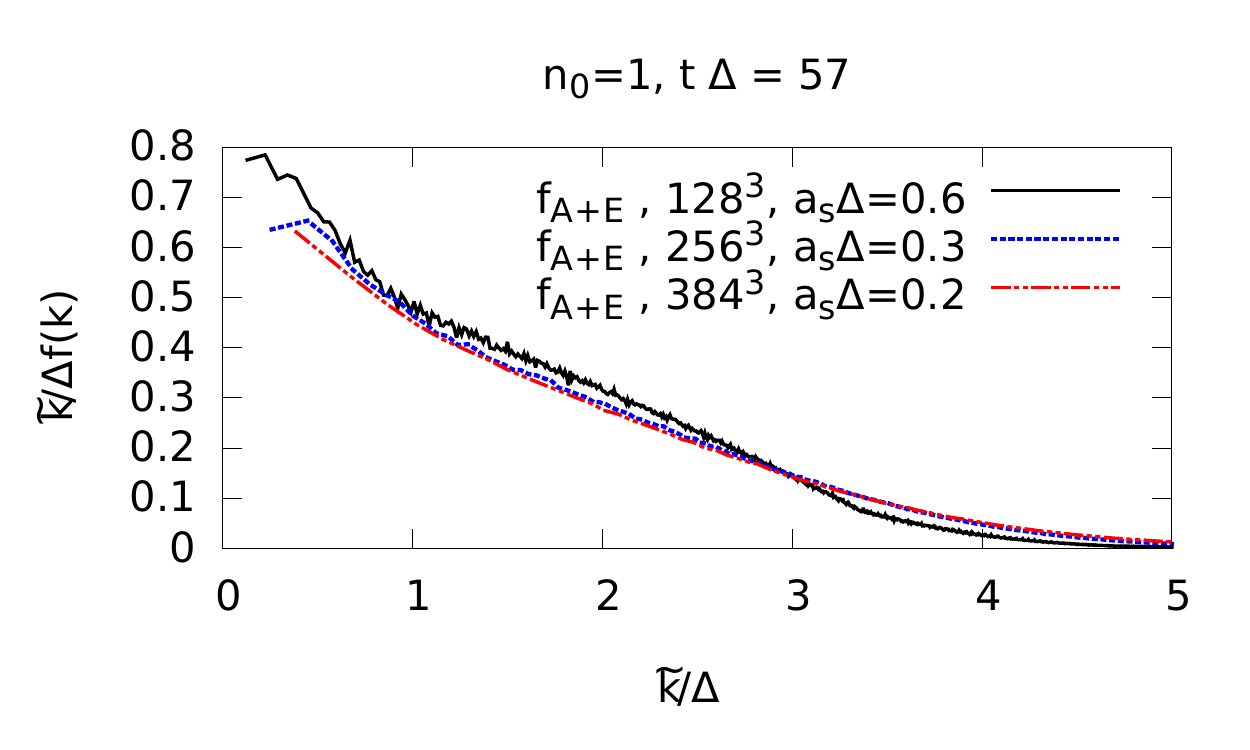}}
 \caption{Integrand of the HTL formula for different values of the UV cutoff with the same IR cutoff, i.e., constant $ L \Delta$.  }
 \label{fig:htlintegrand}
\end{figure}

\subsection{HTL resummed approximation}

We emphasize that in the case of strong occupation numbers of particles at the dominant momentum scale $\Delta$, it is not obvious that a HTL-like separation of scales is a valid picture. If this is the case, however, we should be able to get the plasmon mass from the integral
\begin{equation} \label{eq:htlintegral}
\omega_{pl}^2 = \dfrac{4}{3} g^2 N_c \int \dfrac{\mathrm{d}^3k}{\left(2 \pi \right)^3} \dfrac{f\left(k \right)}{k}.
\end{equation}
On the lattice the integral is discretized by the standard replacement 
\begin{equation}
\int\dfrac{ \mathrm{d}^3k}{\left(2\pi \right)^3} \rightarrow \sum_k\dfrac{1}{V},
\end{equation}
where $k$ runs over the modes available on the lattice. This method is also widely used in the literature, see, e.g., \cite{Epelbaum:2011pc,Berges:2013fga,Mace:2016svc}.
While estimating the mass scale using (\ref{eq:htlintegral}) one can use different definitions for the particle distribution, as discussed in Sec. \ref{sec:partdist}. However, it turns out that this has only a small effect on the values of the mass scale, less than 10\% for the cases considered in this paper.

\begin{figure}[tb!]
\centerline{\includegraphics[width=0.48\textwidth]{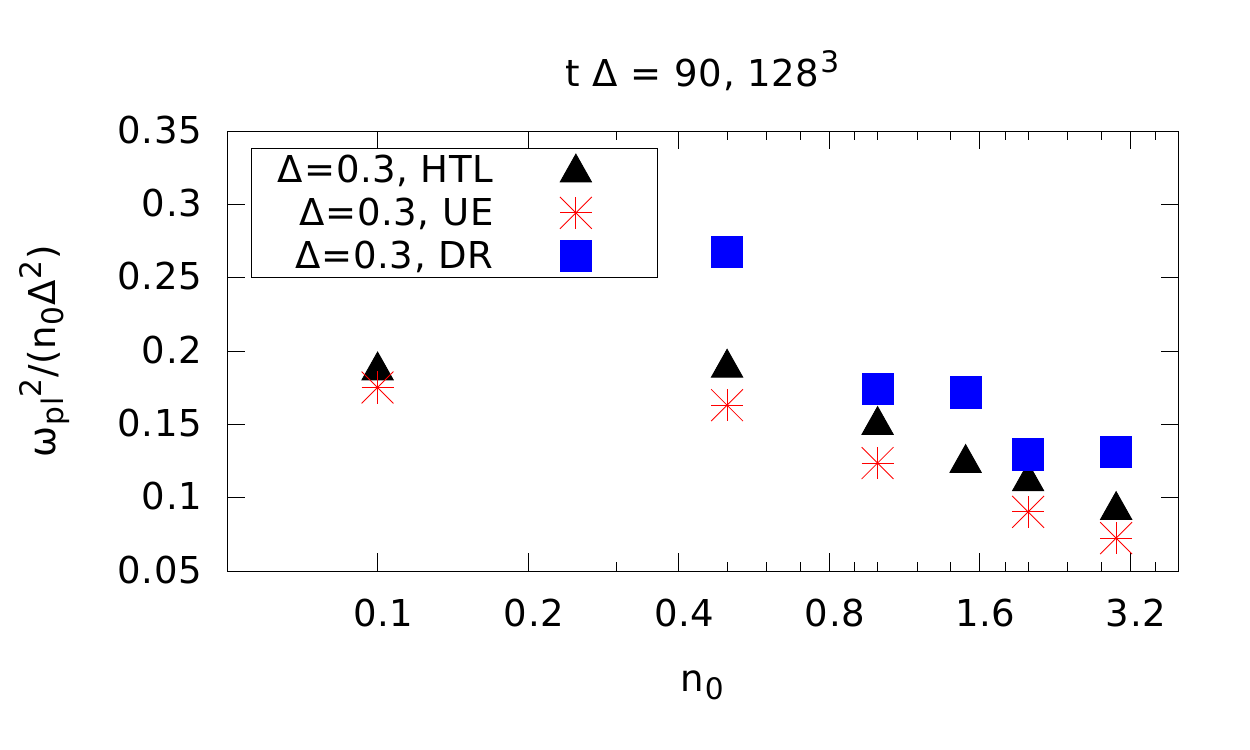}}
 \caption{Dependence of plasmon mass (scaled by the occupation number $n_0$) on the occupation number $n_0$ for the different methods of evaluating the plasmon mass scale. }
\label{fig:occupnumberdep}
\end{figure}

\begin{figure*}[tb!]
\centerline{\includegraphics[width=0.48\textwidth]{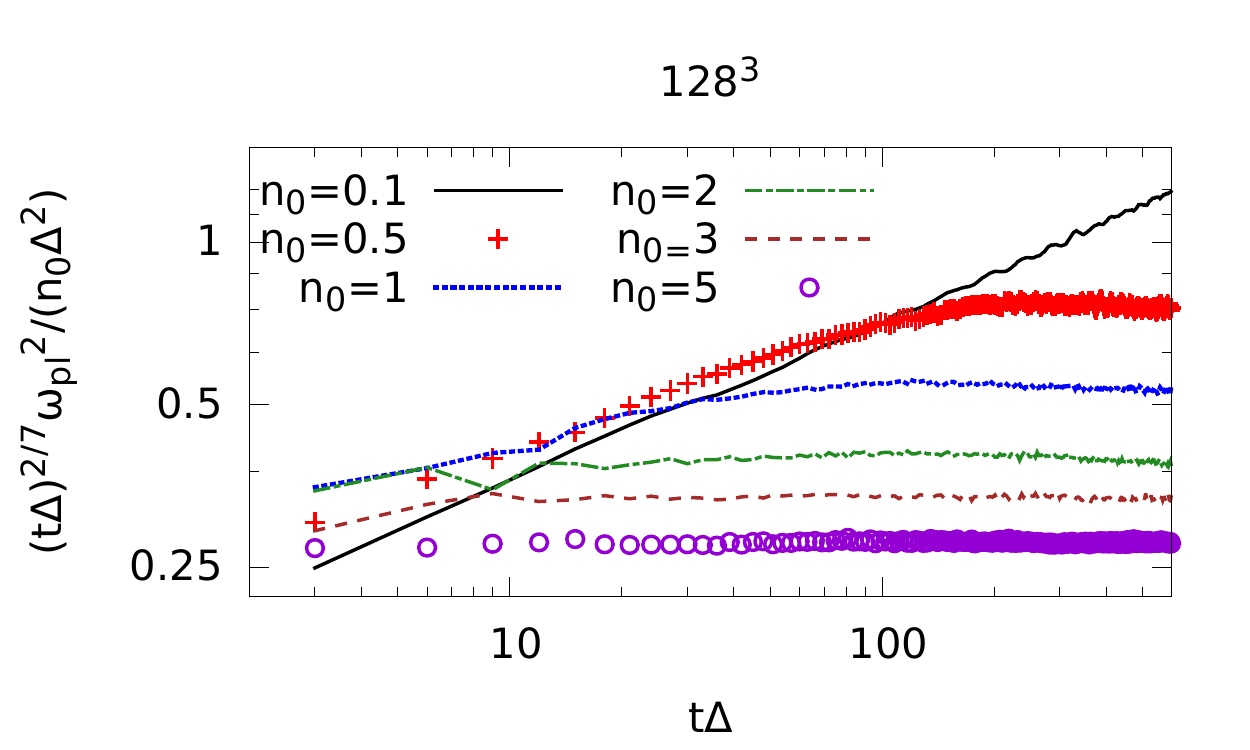}
\includegraphics[width=0.48\textwidth]{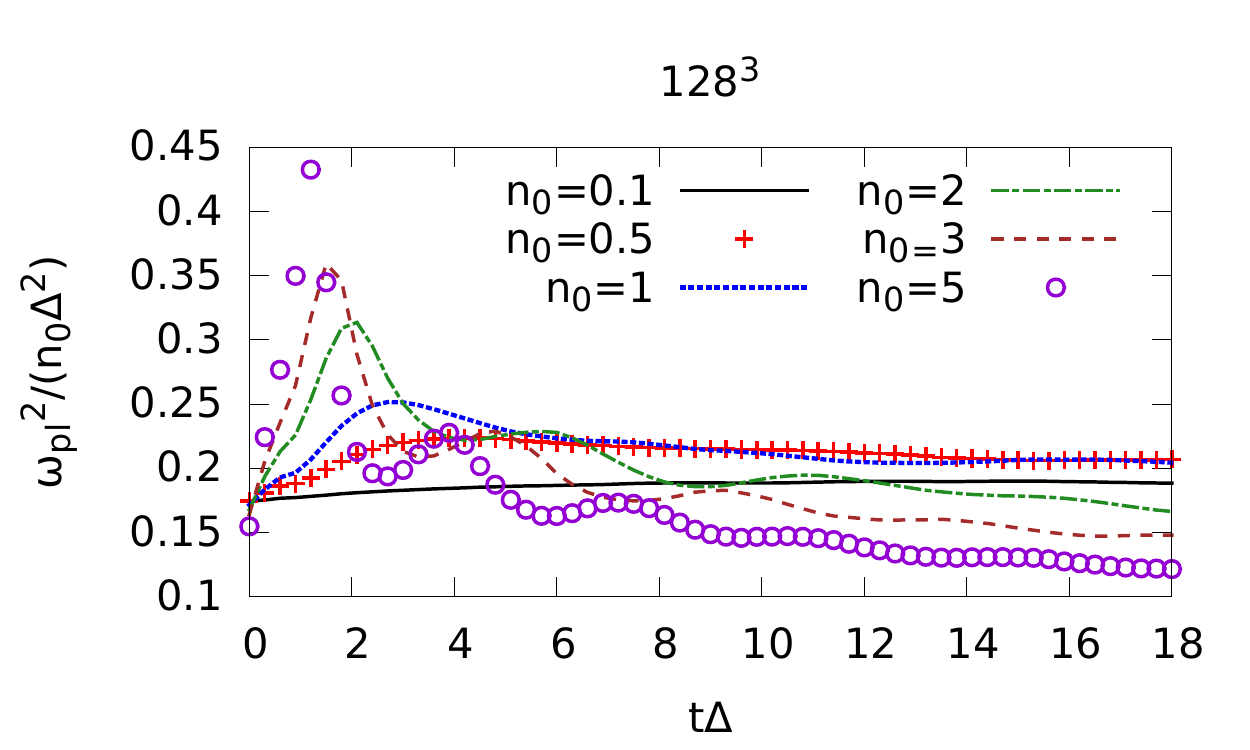}}
 \caption{Time dependence of plasmon mass using the HTL resummed approximation at late times (left) and the initial transient behavior (right). The momentum scale used here was $\Delta=0.3.$  We find that the late-time behavior is consistent with $t^{\nicefrac{-2}{7}}$ power law, as proposed in~\cite{Kurkela:2012hp}, and that the simulations with larger occupation numbers settle faster into this asymptotic behavior. Note that the plasmon mass in the left panel has been scaled by a power of $t \Delta$ which is not the case on the right.}
 \label{fig:tdep}
\end{figure*}

\section{Lattice cutoff dependence} \label{sec:cutoff}

Next we move to study the dependence of the different methods on the IR and UV cutoffs provided by the lattice size $L$ and spacing $a_s$. In the IR, we would expect the number distribution to thermalize to $f(k) \sim 1/k$, yielding a finite plasmon mass at least in the HTL approximation~\nr{eq:htlintegral}. On the other hand, our initial Gaussian occupation number is very suppressed in the UV, and until the system has time to reach a classical thermal equilibrium, we would expect the plasmon mass to be independent of the lattice UV cutoff. On the lattice, the shortest wavelength  we can have i.e., the UV cutoff, is  $\sim a_s$. Correspondingly, the longest wavelength or IR cutoff  is given by $L $. Varying either one of these while keeping the other (and the momentum scale $\Delta$) fixed will reveal us how our results depend on these cutoffs. 

From \figs \ref{fig:ircutoff} and \ref{fig:uvcutoff} we can see the numerical results for the cutoff dependencies. As \fig  \ref{fig:ircutoff}, shows we find no (or very insignificant) infrared cutoff dependence. The HTL resummed approximation and the uniform electric field methods seem to be completely infrared safe. The dispersion relation method does not show a significant IR cutoff dependendence, although its statistical accuracy is worse than that of the other methods.

However, this is not the case for the ultraviolet cutoff, as can be seen from \fig \ref{fig:uvcutoff}. It seems that the hard-thermal-loop resummed approximation has a non-negligible ultraviolet cutoff dependence. Once again the uniform electric field method seems to be completely ultraviolet safe along with the dispersion relation. In order to better understand the behavior of the  HTL approximation, we have studied the dependence of the quasiparticle distribution on the lattice cutoffs. The results can be seen in \fig \ref{fig:htlintegrand}, where the integrand of the HTL formula \eq\nr{eq:htlintegral}, is shown. We see that some non-trivial phase space effect takes place: as we increase the ultraviolet cutoff (which happens when we decrease  $a_s \Delta$), we find a reduction in the occupation numbers in the infrared and an increase in the ultraviolet.   Out of these two effects the dependence of the  occupancy in the IR on the UV cutoff is  the more important one for the value of the integral~\nr{eq:htlintegral}. A possible interpretation of this observation is that the opening up of new UV phase space in the continuum limit allows more energy to be transferred from the IR to the UV, decreasing the occupancy. 
However, it seems that in the continuum limit the HTL resummed approximation approaches the result given by the uniform electric field method, as we can read from \fig \ref{fig:uvcutoff}.

The plasmon masses given by the dispersion relation method are nicely cutoff independent, but are consistently higher that the other two methods by a factor of 50\%. This could point to a very significant fraction of the electric field energy in the IR residing in some nonplasmon modes that do not propagate but are instead damped extremely quickly and are, therefore, not seen in the oscillations of the uniform electric field (the ``Landau cut''). At this stage, however, we do not have a clear interpretation for the surprisingly large difference between the mass gap in the dispersion relation and the other plasmon mass estimators.

\section{Dependence on time and occupation number}\label{sec:results}

We then move to study the dependence of the plasmon mass scale on more physical parameters of the simulation; the initial typical occupation number $n_0$ and time. Figure \ref{fig:occupnumberdep} shows the dependence of the plasmon mass scale on the  initial occupation number at fixed physical time $t\Delta$ (the initial particle  distribution is given by equation (\ref{eq:initdist}) and the parameter $n_0$  determines its overall normalization). The relation between the different methods for  determining $\omega_{pl}$ can be seen  to be independent of $n_0$. The value of $n_0$  controls the strength (or absence) of the scale separation between the plasmon scale  $\omega_{pl}^2 \sim n_0$ and the hard scale $\Delta$. Thus, the validity of the HTL  picture should be regained in the limit $n_0 \to 0$. However, even in the larger $n_0$  results, we see no clear indication of the breakdown of the HTL calculation of the  plasmon mass scale, although the ambiguity related to the different definitions of the  quasiparticle distribution discussed in Sec.~\ref{sec:partdist} grows larger.

We find that at a fixed time the plasmon mass squared increases less than linearly with the initial occupation number $n_0$, while generically a linear dependence would be expected. While this could in principle result from some nontrivial nonlinear effect, the more likely explanation is provided by the different time dependence for different $n_0$.

The dependence of the plasmon mass (obtained using the HTL method) on the scaled time $t\Delta$ is shown in \fig \ref{fig:tdep} in both early and late times. It seems that after initial transient behavior the observed time evolution is qualitatively independent of occupation number, and the plasmon mass scale seems to decrease like a power law. The asymptotic behavior seems to be consistent with $t^{-2/7}$, which was proposed in~\cite{Kurkela:2012hp} based on a kinetic theory analysis of the cascade of energy towards the UV. The duration of the initial transient behavior depends strongly on $n_0$: for large occupation numbers the asymptotic behavior sets in faster. This provides a natural explanation for the less-than linear dependence of the plasmon mass at large $t\Delta$ on the initial occupation $n_0$ observed in Fig.~\ref{fig:occupnumberdep}. For larger $n_0$ the system has spent a larger fraction of its history in the $\omega_{pl}^2 \sim t^{-2/7}$ regime, leading to a smaller $\omega_{pl}^2/n_0$ at a fixed $t\Delta $.

\begin{figure}[tb!]
\centerline{\includegraphics[width=0.48\textwidth]{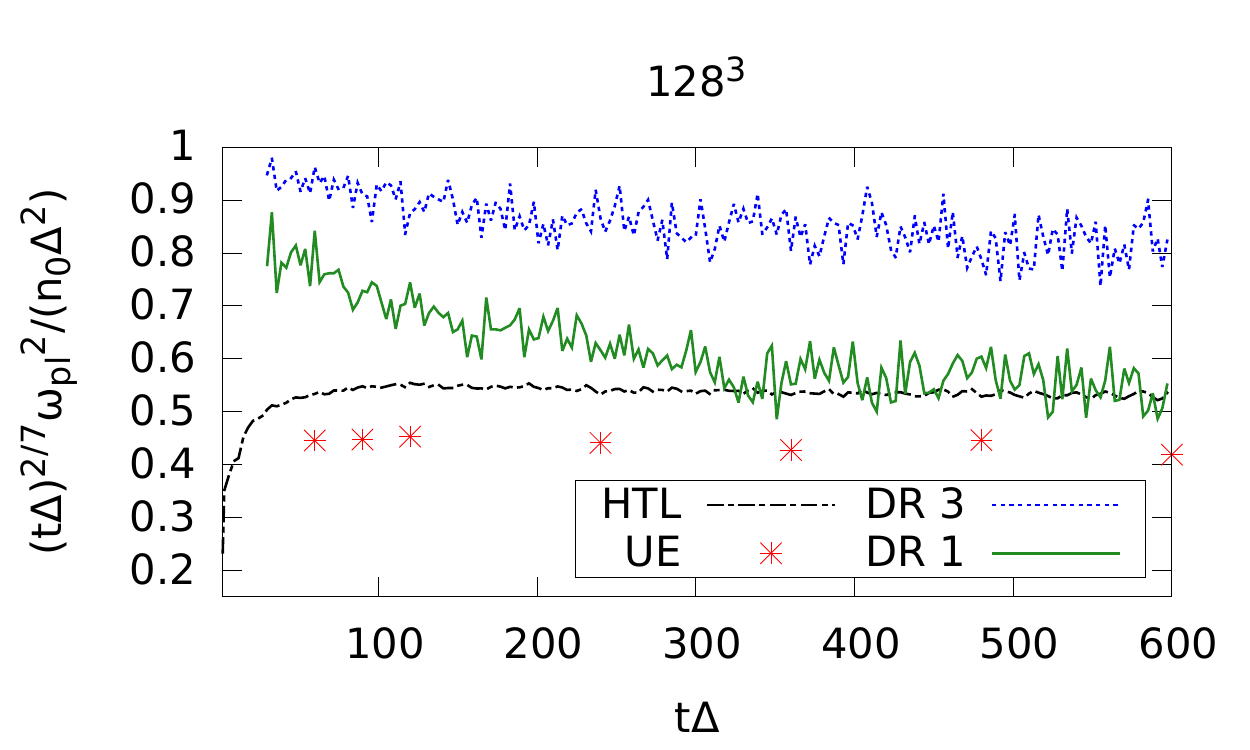}}
 \caption{Comparison of the observed time dependence of plasmon mass using all three methods. For the dispersion relation the numbers 1 and 3 indicate the largest value of  $\nicefrac{k^2}{\Delta^2}$ included in the dispersion relation fit. The momentum scale used here was $\Delta=0.3$ and $n_0=1.$ We find that the late-time behavior is consistent with a $t^{\nicefrac{-2}{7}}$ power law for all methods considering the large uncertainty in the DR method.}
 \label{fig:tdep2}
\end{figure}

We have also studied the time dependence of the plasmon mass using methods other than the HTL resummed approximation. The results are shown in \fig \ref{fig:tdep2}. We find that while the UE and HTL methods are in agreement when it comes to the asymptotic behavior (i.e. the $t^{2/7}$ power law), the result from the DR method suffers from too-large fluctuations and a strong dependence on the details of the fit to make a firm conclusion. The difference between the UE and HTL methods seems to persist, but we expect it to disappear in the continuum limit as was previously observed for a fixed $t\Delta$.
The plot shows the result of the DR method using two different upper limits for the fit range in $k$ used   in the dispersion relation method. The value of the plasmon mass has a strong dependence on this limit. 
The dispersion relation method also requires much more statistics than the other methods. In \fig \ref{fig:tdep2} the results from the UE and HTL methods have been obtained from a single run, but the results for the DR method have been averaged over 20 runs. In spite of this, the statistical fluctuations are still larger than in the HTL method. While drawing firm conclusions from the DR results is thus difficult, it does seem to  agree better with the other methods in the asymptotical time limit, when the $\omega_{pl}^2 \sim t^{-\nicefrac{2}{7}}$ decrease reintroduces a clearer separation between the two mass scales $\omega_{pl}$ and $\Delta$.

\section{Conclusions and outlook}\label{sec:conc}
Based on the results in this paper we consider that the uniform electric field method is the most reliable one to extract the plasmon mass, as it seems to be insensitive to the ultraviolet and infrared cutoff effects. In the future it would be interesting to determine a full dispersion relation $\omega(k)$  with this method by extracting a coordinate-dependent electric field with a specific $k$. However, the better reliability of the UE method comes with a price: it comes with a great computational cost when compared to the other methods, since it requires one to evolve the system for thousands of time steps further in time before one can reliably extract the plasmon mass scale. 

It seems that hard-thermal-Loop approximation can be brought into an agreement with the uniform electric field method when the results are extrapolated to the continuum limit. The continuum extrapolation  is relatively well controlled here since we are working with a very UV-suppressed spectrum of particles. With a power-law spectrum reaching the continuum limit would be more difficult. An important conclusion from the agreement between the UE and HTL methods is that the kinetic theory description in terms of weakly interacting quasiparticles seems indeed to be a valid way to understand the overoccupied classical gauge field system, even quantitatively. This was not obvious a priori, although also earlier numerical studies have pointed in this direction.

The dispersion relation method tends to give larger (of the order of 50 \%) values for the plasmon mass when compared to the other methods, and at present we have no clear interpretation of this difference. The dispersion relation method also requires more statistics to converge to a stable value, while for the UE and HTL methods one can get a very good estimate from a single configuration on the lattice sizes $128^3 \dots 384^3$ used here. The dispersion relation method also has a non-negligible dependence on time and on the details of the fit procedure, which must be interpreted carefully if one wishes to use this method. The disagreement of the dispersion relation method with the two others points to a limitation with the quasiparticle picture in the overoccupied classical system. While the hard modes do generate a plasmon mass scale that can be estimated from the HTL formula, the behavior of the modes at this plasmon scale is more complicated than merely a collection of massive quasiparticles, at least when the separation between the hard and plasmon scales is not large. 

Generalizing our results to the expanding case is not straightforward without performing actual simulations. The reason is that expansion inevitably leads to anisotropy, and instead of one clear momentum scale $\Delta$ we end up having two separate characteristic momentum scales, one in the longitudinal and one in the transverse direction.
An important future development will be to use these same methods to analyze a purely two-dimensional and strongly anisotropic systems closer to the physical situation in a heavy-ion collision. The HTL calculation of the polarization tensor can be extended to the case of an anisotropic momentum distribution of hard modes, but not to a purely two-dimensional system (i.e. the infinitely anisotropic limit). It would be interesting to study these extremely anisotropic systems using this classical field setup, and we plan to return to this in future work.

\begin{acknowledgments}
  We are grateful to K. Boguslavski, A. Kurkela, M. Laine and S. Schlichting  for discussions.
  T.~L.\ is supported by the Academy of Finland, projects No. 267321,
  No. 273464, and No. 303756. J.P. is supported by the Jenny and Antti Wihuri Foundation. The authors wish to acknowledge CSC – IT Center for Science, Finland, for computational resources.
\end{acknowledgments}

\bibliography{spires}
\bibliographystyle{JHEP-2modlong}

\end{document}